\documentclass[letterpaper,titlepage,11pt]{article}
\usepackage{hyperref}

\usepackage{amssymb,amsmath,amsfonts}
\usepackage{epsfig}

\def\clock{{\count0=\time
           \divide\count0 60
           \ifnum\count0<10 0\fi\the\count0
           \multiply\count0 -60 \advance\count0 \time
           :\ifnum\count0<10 0\fi \the\count0
         }}
\newcommand{\timestamp}{{\small\vbox{\hbox{\tt\jobname.tex}
\hbox{\the\day/\the\month/\the\year, \clock}}}}

\newcommand{\be}{\begin{equation}} \newcommand{\ee}{\end{equation}}
\newcommand{\bea}{\begin{eqnarray}} \newcommand{\eea}{\end{eqnarray}}

\newcommand{\CM}{\mathcal{M}}

\newcommand{\id}{\hbox{1\kern-.27em l}}
\newcommand{\sid}{\hbox{\scriptsize1\kern-.27em l}}

\newcommand{\we}{\kern-.1em\wedge\kern-.1em}
\newcommand{\scal}{\kern-.13em\cdot\kern-.13em}

\newcommand{\II}{I\kern-.09em I}

 \newcommand{\C}{\mathbb{C}}

\newcommand{\spa}{\ , \ \ }

\newcommand{\R}{\mathbb{R}}
\newcommand{\T}{\mathbb{T}}

\newcommand{\mt}{\mathfrak{t}}
\newcommand{\ms}{\mathfrak{s}}

\newcommand{\beastar}{\begin{eqnarray*}}
\newcommand{\eeastar}{\end{eqnarray*}}

\newcommand{\ds}{\displaystyle}

\numberwithin{equation}{section}
\begin{document}

\begin{titlepage}

\rightline{\vbox{\small\hbox{\tt hep-th/0510098} }} \vskip 3cm

\centerline{\Large \bf Thermodynamics of the Near-Extremal
NS5-brane} \vskip 1.6cm

\centerline{ {\bf  Troels Harmark} and {\bf Niels A. Obers} }

\vskip 0.5cm
\begin{center}
\sl The Niels Bohr Institute \\
\sl Blegdamsvej 17, 2100 Copenhagen \O , Denmark\vskip 0.5cm
\end{center}

\vskip 0.5cm

\centerline{\small\tt harmark@nbi.dk, obers@nbi.dk}

\vskip 1.6cm

\centerline{\bf Abstract} \vskip 0.2cm \noindent We consider the
thermodynamics of the near-extremal NS5-brane in type IIA string
theory. The central tool we use is to map phases of
six-dimensional Kaluza-Klein black holes to phases of
near-extremal M5-branes with a transverse circle in
eleven-dimensional supergravity. By S-duality these phases
correspond to phases of the near-extremal type IIA NS5-brane. One
of our main results is that in the canonical ensemble the usual
near-extremal NS5-brane background, dual to a uniformly smeared
near-extremal M5-brane, is subdominant to a new background of
near-extremal M5-branes localized on the transverse circle. This
new stable phase has a limiting temperature, which lies above the
Hagedorn temperature of the usual NS5-brane phase. We discuss the
limiting temperature and compare the different behavior of the
NS5-brane in the canonical and microcanonical ensembles. We also
briefly comment on the thermodynamics of near-extremal D$p$-branes
on a transverse circle.

%\vskip 0.5cm \leftline{\timestamp}

\end{titlepage}

%\pagestyle{empty}
%\tableofcontents
%\newpage

\pagestyle{plain} \setcounter{page}{1}
%%%%%%%%%%%%%%%%%%%%%%%%%%%%%%%%%%%%%%%%%%%%%%%%%%%%%%%%%%%%%%

\section{Introduction}

The physics of the NS5-brane is interesting for many reasons: It
is conjectured to have a novel non-gravitational string theory
living on its world-volume
\cite{Seiberg:1997zk,Berkooz:1997cq,Dijkgraaf:1997ku}. The type
IIA NS5-brane world-volume theory has the $(2,0)$ super conformal
field theory as its low energy limit \cite{Seiberg:1997ax}.
Moreover, the NS5-brane wrapped on certain cycles describes pure
super Yang-Mills theory \cite{Maldacena:2000yy}.

In this paper we consider the thermodynamics of the near-extremal
NS5-brane in type IIA string theory. We use as a starting point
that the type IIA NS5-brane is given non-perturbatively as an
M5-brane with a transverse circle. More specifically, we employ
the idea of \cite{Harmark:2002tr,Harmark:2004ws} that in order to
understand the thermodynamic phases of the type IIA NS5-brane it
is necessary to know all possible phases of M5-branes with a
transverse circle.%
\footnote{In this paper we consider only the near-extremal type
IIA NS5-brane, but our results can easily be extended to the full
non-extremal case. However, we restrict ourselves to the
near-extremal case since this has the most interesting physics and
since it is dual to a decoupled non-gravitational theory.}

An important aspect of the NS5-brane is that the thermodynamics of
the usual near-extremal NS5-brane background (dual to an M5-brane
uniformly smeared on a transverse circle) has entropy as function
of energy $S(E) = T_{\rm hg}^{-1} E$, where $T_{\rm hg}$ is a
fixed temperature \cite{Maldacena:1996ya,Maldacena:1997cg}. Thus,
the thermodynamics of the NS5-brane is singular and has
Hagedorn-like behavior, supporting the idea that its
non-gravitational dual is a string theory. In
\cite{Harmark:2000hw,Berkooz:2000mz} the idea was put forward that
the singular thermodynamics of the NS5-brane could be resolved by
computing a string one-loop correction to the background. This was
done in \cite{Kutasov:2000jp}, with the result that the corrected
background has $T > T_{\rm hg}$ and negative specific heat. In a
subsequent analysis the results of \cite{Kutasov:2000jp} have been
interpreted to indicate that one does not have a Hagedorn phase
transition and that the Hagedorn temperature is a limiting
temperature \cite{Aharony:2004xn}.

The main result of this paper is that in the canonical ensemble
the usual near-extremal NS5-brane background is thermodynamically
subdominant to a new eleven-dimensional background of
near-extremal M5-branes localized on the transverse circle. This
means that if one starts with the usual near-extremal NS5-brane
background this will decay and end up in the new phase which is
thermodynamically stable in that it has a positive specific heat.
That the new phase is dominant means in particular that the string
one-loop correction, or any other correction to the usual
near-extremal NS5-brane background, is subdominant to the effect
of the decay to the new eleven-dimensional phase. The new phase is
also dominant in the microcanonical ensemble for sufficiently
small energies.

However, the presence of the new stable phase also raise a puzzle.
The puzzle origins in the fact that the new phases we discuss in
this paper have a maximally possible temperature. Therefore, it is
not clear what happens if one tries to heat up the type IIA
NS5-brane to a temperature higher than this maximal temperature.
As we discuss in the paper, this is also connected to the fact
that the canonical and microcanical ensembles seem to be
inequivalent at large energies.

The new results in this paper on the thermodynamics of the
NS5-brane are obtained using the transformation of
\cite{Harmark:2002tr,Harmark:2004ws} that can take a phase of a
six-dimensional Kaluza-Klein black holes and transform it into a phase
for near-extremal M5-branes with a transverse circle. Our focus in
this paper is on phases without Kaluza-Klein bubbles since these
are the dominant phases in the phase diagram. Phases obtained from
the solutions with Kaluza-Klein bubbles in \cite{Elvang:2004iz} are
considered in \cite{Harmark:2004bb}.

For comparison, we also briefly discuss the thermodynamics of
near-extremal D$p$-branes on a circle. These cases have some
properties in common with the M5-brane case, but also differ
significantly in some other aspects. In particular, they do
neither exhibit a maximal temperature, nor a discrepancy between
the microcanonical and canonical ensembles.

The outline of this paper is as follows. In Section \ref{sec:rev}
we review some of the basic facts on near-extremal NS5-branes that
we build on in the rest of the paper. In Section \ref{sec:KK} we
review how one maps phases of six-dimensional Kaluza-Klein black
holes to phases of near-extremal type IIA NS5-branes, and we use
recently obtained data to find the entire localized phase. In
Section \ref{sec:therm} we discuss the new thermodynamics of the
near-extremal type IIA NS5-brane in the canonical and
microcanonical ensembles. In Section \ref{sec:Dp} we comment on
similarities in the thermodynamics of near-extremal D$p$-branes on
a transverse circle. Finally, we present our conclusions in
Section \ref{sec:con}.

%%%%%%%%%%%%%%%%%%%%%%%%%%%%%%%%%%%%%%%%%%%%%%%%%%%
\section{Basic facts on near-extremal NS5-branes}
\label{sec:rev}

In this section we briefly review basic facts about near-extremal
NS5-branes. Note that the results of
\cite{Harmark:2002tr,Harmark:2004ws} for near-extremal NS5-brane
on which we build in this paper, are not reviewed in
this section but instead in Sections \ref{sec:KK} and
\ref{sec:therm}.

One of the reasons why the thermodynamics of near-extremal
NS5-branes is interesting is the fact that this background is
believed to be dual to a non-gravitational theory. In the
decoupling limit of $N$ coincident NS5-branes in type IIA string
theory the string length $l_s$ is kept fixed while the string
coupling $g_s$ goes to zero
\cite{Maldacena:1997cg,Aharony:1998ub}. In this limit the dynamics
of the theory is believed to reduce to a string theory without
gravity, called Little String Theory (LST), or more precisely
$(2,0)$ LST
of type $A_{N-1}$ \cite{Seiberg:1997zk,Berkooz:1997cq,Dijkgraaf:1997ku}.%
\footnote{See
\cite{Giveon:1999px,Giveon:1999tq,Aharony:1999ks,Narayan:2001dr,DeBoer:2003dd,Aharony:2003vk,Aharony:2004xn,Parnachev:2005hh}
for other interesting work on thermodynamics and correlators of
Little String Theory.}

Consider now instead $N$ non-extremal type IIA NS5-branes. Then
the decoupling limit, or near-extremal limit, is defined as
keeping $l_s$ fixed and taking $g_s$ to zero, while at the same
time keeping the energy above extremality fixed. This gives the
following background (in the string frame)
describing $N$ coincident near-extremal NS5-branes%
\footnote{Note that the radial coordinate $r$ is defined to be
dimensionless in \eqref{UNS5}.}
\begin{equation}
\label{UNS5} \begin{array}{c} \ds ds^2 =  - f dt^2 + \sum_{i=1}^5
dx_i^2  + \frac{N l_s^2}{r^2} f^{-1} dr^2 + N l_s^2 d\Omega_3^2
\spa f = 1 - \frac{r_0^2}{r^2} \ ,
\\
\ds g_s^2 e^{2 \phi} = \frac{N}{r^2}  \spa H_3 = d B_2 = 2 N l_s^2
d \Omega_3 \ .
\end{array}
\end{equation}
Here $\phi$ is the dilaton and $B_2$ is the Kalb-Ramond two-form
potential for the field strength $H_3$ under which NS5-branes are
magnetically charged. The thermodynamics of this background is
\begin{equation}
\label{thermouni} T = T_{\rm hg} \equiv \frac{1}{2\pi \sqrt{N}
l_s} \spa S(E) = T_{\rm hg}^{-1} E \spa \frac{E}{V_5} =
\frac{r_0^2}{ (2\pi)^{5} l_s^{6}} \ .
\end{equation}
This thermodynamics corresponds to the thermodynamic behavior of a
string theory at the Hagedorn temperature, $T_{\rm hg}$ being
the Hagedorn temperature \cite{Maldacena:1996ya,Maldacena:1997cg}.
The NS5-brane description of (2,0) LST is valid at high energies
$E/V_5 \gg N l_s^{-6}$.

The above background corresponds to the exact worldsheet CFT given by
$H_3^+/U(1) \times SU(2)_N \times \R^5$, with $H_3^+ \equiv SL(2,\C)_N/SU(2)_N$,
which can be used to perform string computations in this background.
In particular, the string coupling expansion in the background becomes
an expansion around infinite energy in powers of $1/E$
\cite{Kutasov:2001uf}. Thus, one expects the stringy features of
the world-volume theory of NS5-branes to appear in the high energy
regime $E/V_5 \gg N l_s^{-6}$.

As mentioned in the Introduction it was suggested in
\cite{Harmark:2000hw,Berkooz:2000mz} that one could resolve the
singular thermodynamics of the near-extremal NS5-brane by
computing string corrections to the NS5-brane background. In
\cite{Kutasov:2000jp} this was done using the fact that the
near-extremal NS5-brane background is described by an exact
worldsheet CFT. The result of this computation is that the
corrected background has $T
> T_{\rm hg}$, $F > 0$ and negative specific heat. Subsequently,
the instability of this background was argued in
\cite{Aharony:2004xn} to be a tad-pole instability since the
unstable mode has a potential proportional to $-\log E$, driving
the mode to larger values of the energy. See the conclusions in
Section \ref{sec:con} for further comments on the high energy
regime and comparison to the results of this paper.

Turning to lower energies, we have that using the IIA/M S-duality
the near-extremal NS5-brane can be seen as an M5-brane in M-theory
which is smeared uniformly in a transverse direction. The
background \eqref{UNS5} is therefore equivalent to the following
M-theory background
\begin{equation}
\label{UM5} \begin{array}{c} \ds l_p^{-2} ds^2 =
\frac{r^{2/3}}{N^{1/3} l_s^2} \left[ - f dt^2 + \sum_{i=1}^5
dx_i^2 \right] + \frac{N^{2/3}}{r^{4/3}} \left[ f^{-1} dr^2 + dz^2
+ r^2 d\Omega_3^2 \right] \ ,
\\[4mm] \ds F_4  = dC_3 = 2 N l_p^3 \, dz \, d \Omega_3 \ .
\end{array}
\end{equation}
where $f$ is the same as given in \eqref{UNS5} and $C_3$ is the
three-form potential for the four-form field strength under which
the M5-brane is magnetically charged. This background describes
$N$ coincident near-extremal M5-branes smeared on a transverse
circle.  The transverse circle is parameterized by $z$ which we
take to have period $2\pi$. The background \eqref{UM5} is weakly
curved when $N \gg 1$. The near-extremal limit of non-extremal
M5-branes with a transverse circle is given by taking
$l_p \rightarrow 0$ while rescaling the transverse directions.%
\footnote{The near-extremal limit of the NS5-brane and the one of
the smeared M5-brane are easily related using the S-duality
relations $l_p^3 = g_s l_s^3$ and $R_{\rm 11} = g_s l_s$ where
$R_{\rm 11}$ is the radius of the eleventh direction which in this
case is the transverse circle that the M5-branes are smeared on.}
The thermodynamics is identical to that of the near-extremal
NS5-brane given above, and this description of (2,0) LST is valid
for energies $l_s^{-6} \ll E/V_5 \ll N l_s^{-6}$.

At low energy the background of near-extremal M5-branes smeared on
a transverse direction is unstable to decay to a background of
near-extremal M5-branes localized on the transverse circle
\cite{Harmark:2004ws}. For very low energies the $N$ near-extremal
NS5-branes thus enter a phase which is well-described as $N$
near-extremal M5-branes in eleven dimensional flat space. The dual
theory of this background is the $(2,0)$ super conformal field
theory (SCFT). We shall return to this in the following.

\section{NS5-brane phases from Kaluza-Klein black holes \label{sec:KK}}

In this section we briefly review the map of
\cite{Harmark:2002tr,Harmark:2004ws}%
\footnote{See \cite{Harmark:2003fz,Harmark:2005pq} for reviews.}
by which one can obtain solutions for near-extremal M5-branes with
a transverse circle from phases of Kaluza-Klein black holes in six
dimensions. By six-dimensional Kaluza-Klein black holes we mean
pure gravity solutions of six-dimensional General Relativity that
at asymptotic infinity have the geometry of Kaluza-Klein
space-time $\CM^5 \times S^1$, $\CM^5$ being five-dimensional
Minkowski space (see the reviews \cite{Harmark:2005pp,Kol:2004ww}
for more on Kaluza-Klein black holes). As we discuss below, the
near-extremal M5-brane phases can be seen as phases of
near-extremal type IIA NS5-branes.

As described in \cite{Elvang:2004iz,Harmark:2005pp} there are two
classes of Kaluza-Klein black holes in six dimensions: Solutions
with a local $S0(4)$ symmetry, and solutions without a local
$SO(4)$ symmetry. The latter we shall briefly discuss in the
Conclusions, and these are ones that contain Kaluza-Klein bubbles.
For the former class of solutions, which do have a local $SO(4)$
symmetry and which are the ones of interest in this paper, it has
been proven \cite{Wiseman:2002ti,Harmark:2003eg} that the metric
can be put in the ansatz \cite{Harmark:2002tr}
\begin{equation}
\label{ansatz} ds^2 = - f dt^2 + \frac{A}{f} dR^2 +
\frac{A}{K^{3}} dv^2 + K R^2 d\Omega_3^2 \spa f =
1-\frac{R_0^2}{R^2} \ .
\end{equation}
Here $A(R,v)$ and $K(r,v)$ are two functions determining the
metric. $R \rightarrow \infty$ is the asymptotic region where $A$
and $K$ go to one, while $R=R_0$ is the event horizon
\cite{Harmark:2002tr}. We have $v \equiv v + 2\pi$, i.e. we take
the asymptotic circle to have unit radius for simplicity.

Using the map found in \cite{Harmark:2002tr,Harmark:2004ws} we can
now take any six-dimensional Kaluza-Klein black hole solution of
the form \eqref{ansatz} and transform it to a solution of eleven
dimensional supergravity for $N$ near-extremal M5-branes with a
transverse circle given by \cite{Harmark:2002tr,Harmark:2004ws}
\begin{equation}
\label{M5} \begin{array}{c} \ds l_p^{-2} ds^2 =
\frac{R^{2/3}}{N^{1/3} l_s^2} \left[ - f dt^2 + \sum_{i=1}^5
dx_i^2 \right] + \frac{N^{2/3}}{R^{4/3}} \left[ \frac{A}{f} dR^2 +
\frac{A}{K^{3}} dv^2 + K R^2 d\Omega_3^2 \right] \ ,
\\[4mm] \ds F_4  = dC_3 = 2 N l_p^3 \, dz \, d \Omega_3 \ .
\end{array}
\end{equation}
Here $f$ is as in \eqref{ansatz}. The supergravity solution
\eqref{M5} is valid as long as $N \gg 1$, since this guaranties
the geometry to be weakly curved. Note that in the near-extremal
limit $l_p \rightarrow 0$ while $l_s$ is finite. The relation
between the Planck length $l_p$ and the string length $l_s$ is
explained in the previous section. See
\cite{Harmark:2002tr,Harmark:2004ws} for more on the near-extremal
limit of M5-branes with a transverse circle.

The important point is now that {\sl the type IIA NS5-brane is
non-perturbatively defined as an M5-brane with a transverse
circle}. Therefore, the phases of $N$ coincident near-extremal
type IIA NS5-branes are in fact the phases of $N$ coincident
M5-branes with a transverse circle. We can thus get insight into
the thermodynamics of near-extremal NS5-branes by examining
near-extremal M5-branes with a transverse circle as obtained from
phases of six-dimensional Kaluza-Klein black hole solutions.

The relevant physical quantities for the near-extremal M5-brane
solutions with a transverse circle of the form \eqref{M5} are
given as
\cite{Harmark:2002tr,Harmark:2004ws}%
\footnote{Note that these relations imply the Smarr formula $5 TS
= 2 (3-r ) E$ \cite{Harmark:2004ws}.}
\begin{equation}
\label{M5quant}
E = \frac{V_5 R_0^2(1-\chi)}{(2\pi)^5 l_s^6}  \spa r =
\frac{1-6\chi}{2-2\chi} \spa T = \frac{1}{2\pi \sqrt{N} l_s
\sqrt{A_h}} \spa S = \frac{V_5 \sqrt{N} R_0^2 \sqrt{A_h}}{(2\pi)^4
l_s^5 } \ ,
\end{equation}
where $E$ is the energy, $r$ is the relative tension (see
\cite{Harmark:2004ch,Harmark:2004ws} for definition), $T$ is the
temperature and $S$ is the entropy. We define here the number
$\chi$ for a given solution by $K(R,v) = 1 - \chi R_0^2 /R^2 +
\cdots$ in the asymptotic region $R \rightarrow \infty$, and $A_h$
is the value of $A$ on the horizon: $A_h = A(R,v) |_{R=R_0}$. We
can then relate the physical quantities \eqref{M5quant} for
near-extremal M5-branes with a transverse circle to the physical
quantities of the corresponding six-dimensional Kaluza-Klein black
hole by the map \cite{Harmark:2004ws}
\begin{equation}
\label{map} \frac{E}{V_5} = \frac{5+n}{8} \frac{\mu}{ (2\pi)^5
l_s^6} \spa r = \frac{8n}{5+n} \spa T = T_{\rm hg} \mt \sqrt{\mt
\ms} \spa \frac{S}{V_5} = \frac{1}{(2\pi)^5 l_s^6 T_{\rm hg}}
\frac{\ms}{ \sqrt{\mt \ms}} \ .
\end{equation}
Here $\mu$ is the rescaled dimensionless mass, $n$ is the relative
tension, $\mt$ is the rescaled dimensionless temperature and $\ms$ is
the rescaled entropy of the six-dimensional Kaluza-Klein black hole,
as defined in \cite{Harmark:2003dg,Harmark:2005pp}.%
\footnote{For six-dimensional Kaluza-Klein black hole solutions
with metric of the form \eqref{ansatz} we have
\cite{Harmark:2002tr,Harmark:2005pp}
$$ \mu =  R_0^2
\left[ \frac{3}{2} - \chi \right] \spa n = \frac{1- 6\chi}{3-2\chi}
\spa \mt =  \frac{1}{\sqrt{A_h} R_0} \spa \ms = \sqrt{A_h} R_0^3 \ .
$$
}

\begin{figure}[ht]
\centerline{\epsfig{file=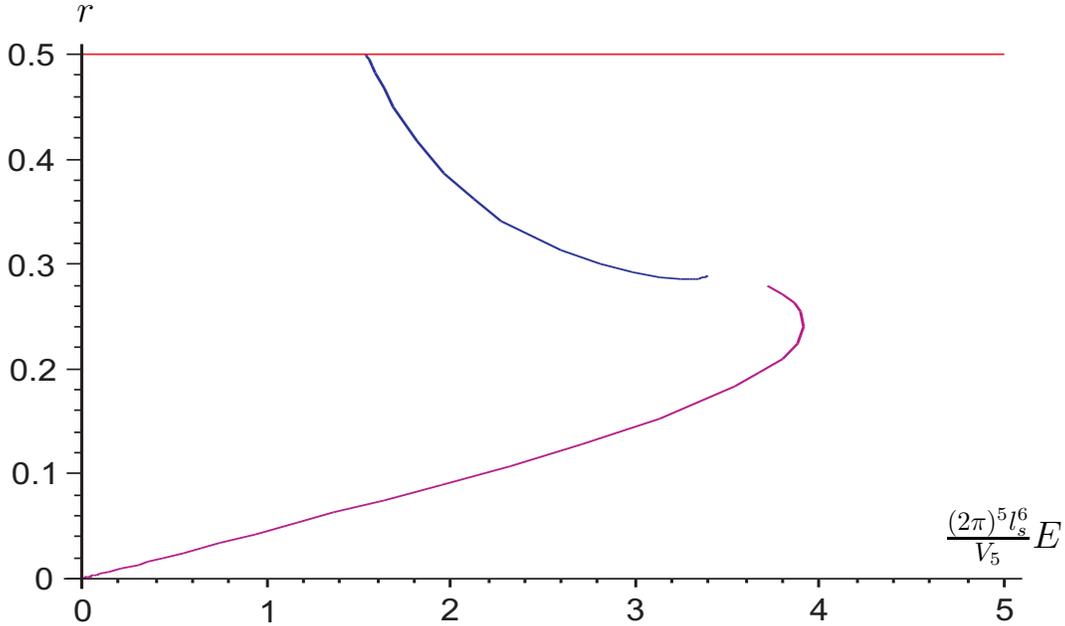,width=14cm,height=8cm} }
\caption{Relative tension versus energy for the uniform (red),
non-uniform (blue) and localized (magenta) phase of near-extremal
NS5-branes. \label{figm5a}}
\begin{picture}(0,0)(0,0)
\put(47,290){\Large $r$} \put(375,91){\Large $\frac{(2\pi)^5
l_s^6}{V_5} E$}
\end{picture}
\end{figure}

We now consider the solutions for $N$ near-extremal M5-branes with
a transverse circle obtained from transforming six-dimensional
Kaluza-Klein black hole solutions with local $SO(4)$ symmetry, via
the map from \eqref{ansatz} to \eqref{M5}. The Kaluza-Klein black
holes with local $SO(4)$ symmetry come in three branches: The
uniform phase, the non-uniform phase and the localized phase. We
therefore naturally obtain the following three phases for $N$
near-extremal type IIA NS5-brane, i.e. for $N$ near-extremal
M5-branes with a transverse circle \cite{Harmark:2004ws}:
\begin{itemize}
\item {\bf Uniform phase.} This phase corresponds to $N$
near-extremal M5-branes uniformly smeared on a transverse circle.
This is the background given by \eqref{UM5}. As discussed in
Section \ref{sec:rev} this background corresponds to the usual
near-extremal NS5-brane background given by \eqref{UNS5}. The
relative tension is $r= 1/2$ for the uniform phase. Note that from
the six-dimensional Kaluza-Klein black hole point of view the uniform
phase corresponds to the uniform black string phase.
\item {\bf Non-uniform phase.} The non-uniform phase corresponds
to the new phase discussed in Ref.~\cite{Harmark:2004ws} of
near-extremal M5-branes non-uniformly distributed on the
transverse circle. The non-uniform phase emanates from the uniform
phase at the critical energy%
\footnote{Note that the critical energy
is obtained from the neutral Gregory-Laflamme mass using the map
\eqref{map}.}
 $E_{\rm c}/V_5 =  1.54 \cdot
(2\pi)^{-5} l_s^{-6} $. The slope in the $(E,r)$ diagram (Figure
\ref{figm5a}) at this point is given by $r(E) \simeq 1/2 - 0.39
\cdot (2\pi)^5 l_s^6 (E-E_{\rm c})/V_5$ \cite{Harmark:2004ws}.
Note that the non-uniform phase is mapped from the neutral
non-uniform black string phase in six dimensions. The solutions
for the neutral non-uniform black string were obtained numerically
in \cite{Wiseman:2002zc} and the behavior near the
Gregory-Laflamme mass was found in \cite{Sorkin:2004qq}.
\item {\bf Localized phase.} The localized phase corresponds to a
background of near-extremal M5-branes localized on the transverse
circle. For $E \rightarrow 0$ we have that $r \rightarrow 0$
corresponding to the fact that the background is well-described by
near-extremal M5-branes in eleven dimensional flat space (this
background is dual to $(2,0)$ SCFT). The first correction to the
background due to the presence of the transverse circle was
obtained in Ref.~\cite{Harmark:2004ws} using the analytical
results on neutral small black holes on cylinders of
\cite{Harmark:2003yz} (see also
\cite{Gorbonos:2004uc,Gorbonos:2005px}). The slope of the curve
for the localized phase at the origin is given by $r(E) \simeq [ 9
\zeta(3)/(25 \pi^2) ] (2\pi)^5 l_s^6 E/V_5$ \cite{Harmark:2004ws}.
Beyond the analytical results for small energies, the localized
phase is known numerically by mapping the numerical solutions
 for localized black holes in six-dimensional
Kaluza-Klein space obtained in Ref.~\cite{Kudoh:2004hs}. These
newly obtained data is in fact what enables us in this paper to
gain new insights into the thermodynamics of the NS5-brane, as we
shall see in Section \ref{sec:therm}.
\end{itemize}
We have depicted the three phases%
\footnote{Note that for the non-uniform and localized phase there
are extra phases in the form of copies since the solutions can be
copied $k$ times on the circle
\cite{Horowitz:2002dc,Harmark:2003eg,Harmark:2004ws}.
 At the level of solutions this means that given the solution \eqref{M5}
  we obtain for any $k=2,3,...$
a new solution with $A'(R,v)= A(kR,kv)$, $K'(R,v) = K (k R, kv)$
and $R_0' = R_0/k$. This solution will have the physical quantities
$E' = E/k^2$, $ r' =r$.}
 in an $(E,r)$ diagram in Figure
\ref{figm5a}. Note that part of the phase diagram for the
near-extremal type IIA NS5-brane was obtained in
\cite{Harmark:2004ws}, but here we add the extra numerical data of
\cite{Kudoh:2004hs}.

The uniform and non-uniform phase have horizon topology $R^5
\times S^3 \times S^1$, while the localized phase has $R^5 \times
S^4$. The $(E,r)$ diagram in Fig.~\ref{figm5a} suggests that the
non-uniform and localized phase meet each other in a topology
changing transition. In fact, from the neutral Kaluza-Klein black
hole point of view, Ref.~\cite{Kudoh:2004hs} presented compelling
evidence that the two branches indeed are meeting by examining
various physical quantities of the numerical solutions.

The uniform phase of near-extremal NS5-branes is classically
unstable for energies below the critical energy $E_{\rm c}$
\cite{Bostock:2004mg,Aharony:2004ig,Harmark:2004ws,Kudoh:2005hf,Harmark:2005jk}.
In this regime the background is expected to decay to the
localized phase corresponding to near-extremal M5-branes localized
on a circle. Seen from type IIA String Theory this is a
non-perturbative phenomenon since it involves localization on the
M-theory circle.

It is important to note that the non-uniform and
localized phases have non-singular event horizons. This one can see
from i) the fact that the map from \eqref{ansatz} to \eqref{M5} takes
a neutral solution with  non-singular horizon to a near-extremal
solution with non-singular horizon and ii) that it is known from
the numerical results of \cite{Wiseman:2002zc,Kudoh:2004hs} that the neutral non-uniform black string
and localized black hole phases have non-singular horizon.
This can also be seen from \eqref{M5} directly, by noting that the
horizon is non-singular if $A(R,v)$ and $K(R,v)$ are well-behaved
for $R \rightarrow R_0$.

%%%%%%%%%%%%%%%%%%%%%%%%%%%%%%%%%%%%%%%%%%%%%%%%%%%%%%%%
\section{New thermodynamics of the NS5-brane}
\label{sec:therm}

In this section we describe the thermodynamics of the
near-extremal NS5-brane using the input of the new phases obtained
from six-dimensional Kaluza-Klein black holes, as discussed in
Section \ref{sec:KK}.

\subsubsection*{Canonical ensemble}

\begin{figure}[ht]
\centerline{\epsfig{file=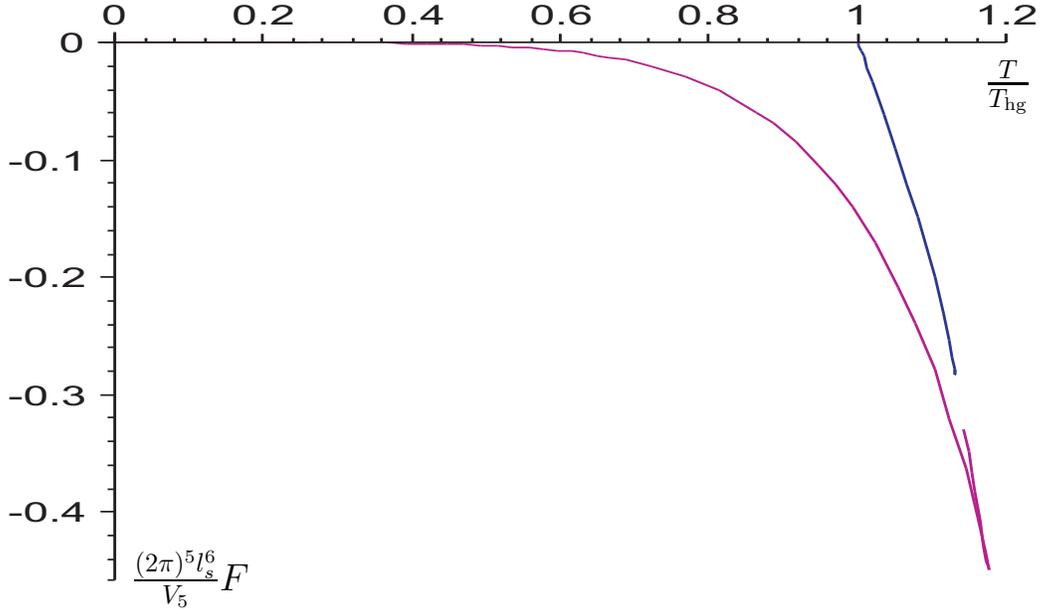,width=14cm,height=8cm} }
\caption{Free energy versus temperature for the near-extremal
NS5-branes. Displayed are the non-uniform (blue) and localized
(magenta) phases. The uniform phase is located in the point
$(T,F)=(T_{\rm hg},0)$. \label{figm5temp} }
\begin{picture}(0,0)(0,0)
\put(388,263){\Large $\frac{T}{T_{\rm hg}}$} \put(65,75){\Large
$\frac{(2\pi)^5 l_s^6}{V_5} F$}
\end{picture}
\end{figure}

We begin with the canonical ensemble since this is the one with
the most interesting results. We have displayed the free energy
versus the temperature in Fig.~\ref{figm5temp}.

Before considering the new data for the localized phase, we first
list some important properties of the three different phases
(already stated in \cite{Harmark:2004ws}):%
\footnote{Note that the non-uniform and localized phase have copies
with free energy  $F' = F/k^2$ and temperature $T' = T$ ($k=2,3,...$).
We do not consider the copied phases here since they are subdominant in
this ensemble.}
\begin{itemize}
\item {\bf Uniform phase.} This phase corresponds to zero free
energy and a fixed temperature $T = T_{\rm hg}$ (as reviewed in
Section \ref{sec:rev}). It corresponds therefore to a single point
in the free energy versus temperature diagram in Figure
\ref{figm5temp}.
\item {\bf Non-uniform phase.} This phase starts out in point
$(T,F) = (T_{\rm hg},0)$ and then has increasing temperature and
decreasing free energy. It has positive specific heat. For
temperatures near the Hagedorn temperature $0 \leq T-T_{\rm hg}
\ll T_{\rm hg}$, we have \cite{Harmark:2004ws}
\begin{equation}
F_{\rm nu} (T ) \simeq - 2 \pi V_5 N^3  T_{\rm hg}^6 \left[ 1.54
\cdot \left( \frac{ T}{ T_{\rm hg}} - 1 \right) + 3.23 \cdot
\left( \frac{ T}{ T_{\rm hg}} - 1 \right)^2 \right] \ .
\end{equation}
This was computed using results of \cite{Sorkin:2004qq}.
\item
{\bf Localized phase.} The localized phase starts in the point
$(T,F)=(0,0)$ which corresponds to the $(2,0)$ SCFT, being an
infrared fixed point of the type IIA near-extremal NS5-brane. The
first correction to the thermodynamics when moving away from the
infrared fixed point by raising the temperature is
 \cite{Harmark:2004ws}
\begin{equation}
F_{\rm loc} (T) = - \frac{2^6 \pi^3}{3^7} V_5 N^{3}  T^6 \left[ 1
+
 \frac{2^5 \zeta (3)}{3^6}
 \frac{T^6}{T_{\rm hg}^6} + {\cal{O}}\left( \frac{ T^{12}}{T_{\rm hg}^{12}}
  \right) \right] \ .
\end{equation}
This was computed using results of \cite{Harmark:2003yz}. We
describe further the thermodynamics of this phase below.
\end{itemize}

We consider now the localized phase, using the data of
\cite{Kudoh:2004hs}. The localized phase starts in the point
$(T,F)=(0,0)$ and has decreasing free energy until the maximum
temperature $T_{\rm \star} = 1.177 \cdot T_{\rm hg}$ is reached.
Then the curve reverses direction and has decreasing temperature
and increasing free energy.

\begin{figure}[ht]
\centerline{\epsfig{file=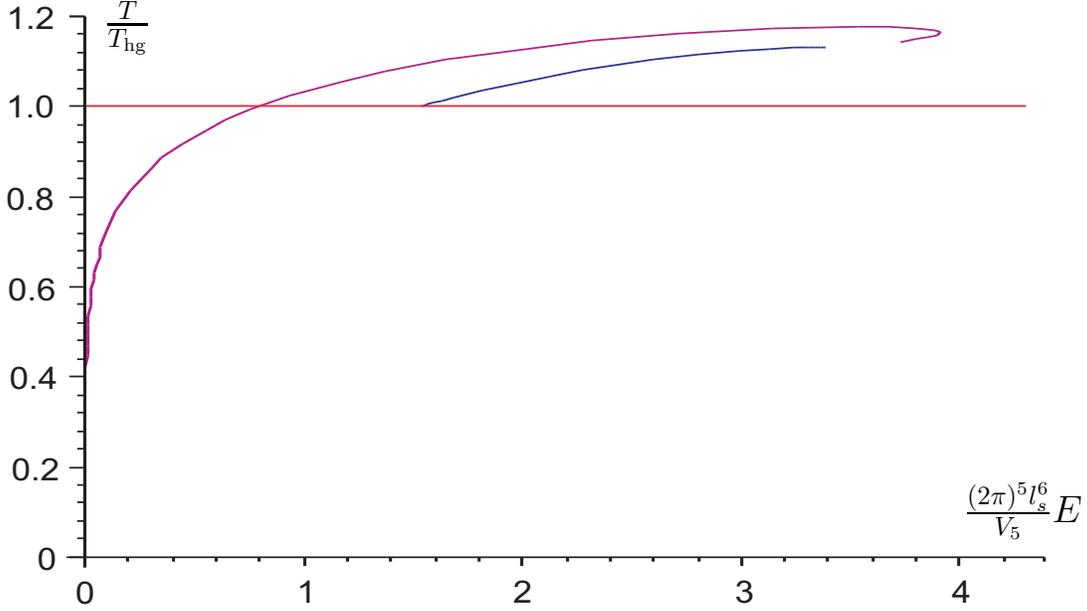,width=14cm,height=8cm} }
\caption{Temperature versus energy for the near-extremal
NS5-branes. Displayed are the uniform (red), non-uniform (blue)
and localized (magenta) phases. \label{figm5te} }
\begin{picture}(0,0)(0,0)
\put(50,275){\Large $\frac{T}{T_{\rm hg}}$} \put(375,90){\Large
$\frac{(2\pi)^5 l_s^6}{V_5} E$}
\end{picture}
\end{figure}

The most interesting part of the localized phase is evidently the
behavior near the maximal temperature $T_{\rm \star}$. A way to
understand the behavior around this point is to consider the
temperature versus energy diagram in Fig.~\ref{figm5te}. The
localized phase in the $(E,T)$ diagram is evidently a smooth
curve. Therefore, the entropy is continuous for $T \simeq T_{\rm
\star}$. However, since the temperature is maximal for $T= T_{\rm
\star}$ the $(E,T)$ curve has a horizontal tangent in that point.
In terms of the specific heat this means that the localized phase has
positive specific heat as $T$ increases from zero to $T_{\rm
\star}$, but then reaches infinite specific heat at $T= T_{\rm
\star}$. If one continue along the curve in the $(E,T)$ diagram we
start from minus infinite specific heat and continue with negative
specific heat until the curve has a vertical tangent where the specific
heat is zero. Then the specific heat is positive for the remaining
part of the curve.%
\footnote{In terms of the $(E,r)$ phase diagram the point with
infinite specific heat is characterized by the $r$ value for
which $E r'(E) = 1/2-r(E)$, while zero specific heat occurs when
$r'(E)$ is infinite. It is not difficult to see from the phase diagram
in Fig.~\ref{figm5a} that for each of these two equations there exists one
particular  solution for $r$ on the localized branch.}

It is interesting to consider more closely the thermodynamics around
the critical temperature $T_{\star}$. From the data we get
\begin{equation}
T \simeq T_{\star} - c (E - E_{\star} )^2  \spa c = 0.038 \spa
\frac{E_{\star}}{V_5} = 3.60 \cdot (2\pi )^{-5} l_s^{-6} \ .
\end{equation}
From this we can obtain the free energy as a function of temperature
near the singularity as
\begin{equation}
\begin{array}{c} \ds
F (T) \simeq F_{\star} - S_{\star} (T - T_{\star}) \pm \frac{2}{3
\sqrt{c} T_{\star}} (T- T_{\star})^{3/2} \ ,
\\[3mm] \ds
\frac{S_{\star}}{V_5} = 3.5 \cdot (2\pi)^{-4} \sqrt{N} \spa
 \frac{F_{\star}}{V_5} \equiv \frac{1}{V_5}(E_{\star} - S_{\star} T_{\star}) =
  - 0.53 \cdot (2\pi)^{-5} l_s^{-6} \ .
\end{array}
\end{equation}
Here the $\pm$ sign corresponds to the two branches on either side of the
maximum temperature critical point. In particular, the minus branch is the one that has positive
specific heat (so with $ E \lesssim E_{\star}$) and the plus branch is the one
that has negative specific heat (so with $E \gtrsim E_{\star}$).
From this expression it is immediately clear that indeed the entropy is finite
and continuous at the critical point, while the heat capacity diverges and
changes sign.

From Fig.~\ref{figm5temp} it is clear that there are temperatures where more than
one phase is available. To understand what happens for these temperatures,
we should first recall that in order for the supergravity
solution to be weakly curved we need $N \rightarrow \infty$.
Moreover, the free energy for a particular supergravity solution will
diverge as $N \rightarrow \infty$. Considering now two phases
at the same temperature with
free energies $F_1$ and $F_2$ where $F_2 > F_1$, we can see that
the phase with the lowest free energy dominates the Euclidean path integral,
since we have that
$ Z \simeq e^{ -\beta F_1} + e^{ -\beta F_2}
= e^{-\beta F_1} ( 1 + e^{-(F_2 - F_1)}) \simeq e^{-\beta F_1} $.

Considering then the phase diagram of Fig.~\ref{figm5temp}, we see
that we have two phases available when $ T_{\rm hg} \leq T \leq
T_{\star}$, but we can conclude from the above that the phase with
lowest free energy dominates. We notice that the dominating phase
corresponds to the localized phase in the energy range $0 \leq E
\leq E_{\star}$. This means for example that if one starts in the
uniform phase i.e. with zero free energy and temperature $T=T_{\rm
hg}$, then, in the canonical ensemble, the system will decay to
the localized solution at the same temperature.

Imagining now that we are in the dominating phase, which is the localized
phase as described above, we can consider what happens as we increase the
temperature. The heat capacity will increase and eventually becomes infinite
once we reach the temperature $T=T_{\rm \star}$. This means that $T_{\star}$
is a limiting temperature. Note that this limiting temperature
exceeds the temperature $T_{\rm hg}$ associated with the uniform phase.

%%%%%%%%%%%%%%%%%%%%%%%
\subsubsection*{Microcanonical ensemble}

\begin{figure}[ht]
\centerline{\epsfig{file=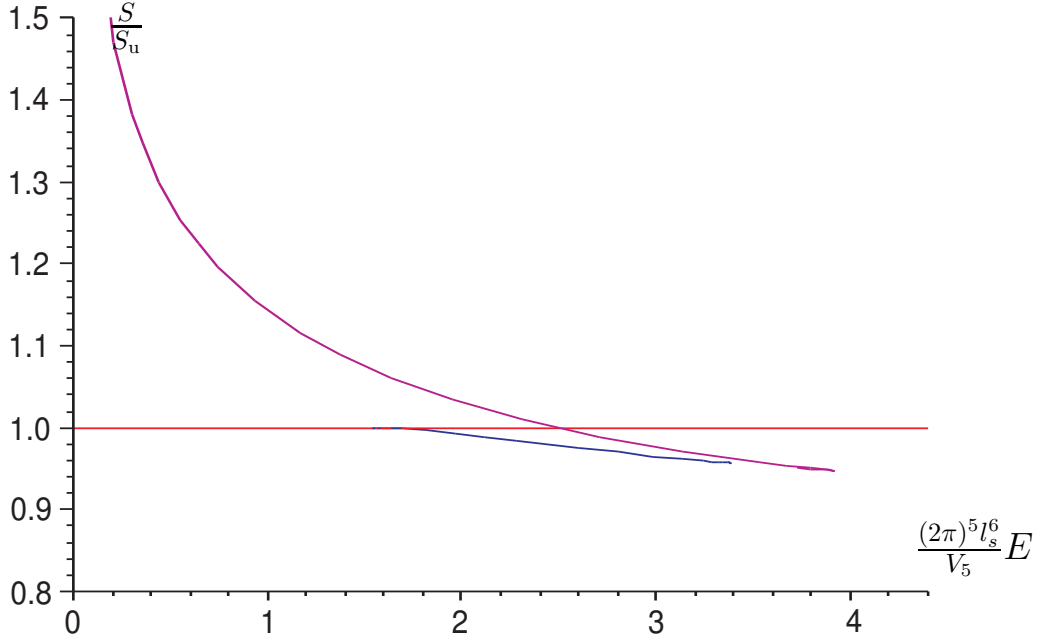,width=13cm,height=9cm} }
\caption{Entropy versus energy for the near-extremal NS5-branes.
On the vertical axis we have plotted $S(E)/S_{\rm u}(E)$ and on
the horizontal axis $E/E_{\rm c}$. Here $S_{\rm u}(E)$ is the
entropy function for the uniform phase of near-extremal
NS5-branes. Displayed are the uniform (red), non-uniform (blue)
and localized (magenta) phases.  \label{figm5ent} }
\begin{picture}(0,0)(0,0)
\put(73,328){\Large $\frac{S}{S_{\rm u}}$} \put(378,130){\Large
$\frac{(2\pi)^5 l_s^6}{V_5} E$}
\end{picture}
\end{figure}

We now turn to the microcanonical ensemble. In Fig.~\ref{figm5ent}
we have displayed the entropy of the three different phases
normalized in terms of the entropy of the uniform
phase as a function of energy.
The three different phases have the following properties:%
\footnote{Note that the non-uniform and localized phase have copies
with entropy  $S' = S/k^2$ and energy $E' = E/k^2$ ($k=2,3,...$).
We do not consider the copied phases here since they are subdominant in
this ensemble.}

\begin{itemize}
\item {\bf Uniform phase.}
This phase is at the Hagedorn temperature, so the entropy (see \eqref{thermouni})
is given by
\begin{equation}
S_{\rm u} (E) = \frac{E}{T_{\rm hg}} \ ,
\end{equation}
and corresponds therefore to the horizontal line in Fig.~\ref{figm5ent}.
\item {\bf Non-uniform phase.}
This phase starts out in the point $(E,S)=(E_{\rm c},S_{\rm c})$,
 and for energies close to the critical energy $E_{\rm c}$
we have \cite{Harmark:2004ws}
\begin{equation}
\frac{S (E)}{S_u (E)} \simeq 1 - 0.05 \cdot \left[ \frac{(2\pi)^5
l_s^6}{V_5} (E- E_{\rm c}) \right]^2 \spa \frac{E_{\rm c}}{V_5} =
1.54 \cdot (2\pi)^{-5} l_s^{-6}  \spa 0 \leq \frac{l_s^6
(E-E_c)}{V_5} \ll 1 \ .
\end{equation}
This was computed using results of \cite{Sorkin:2004qq}. The
entropy function for the entire non-uniform phase, as displayed in
Fig.~\ref{figm5ent}, was computed in \cite{Harmark:2004ws} using
the data of \cite{Wiseman:2002zc}.
\item {\bf Localized phase.} The localized phase starts out in the
point $(E,S)=(0,0)$. At very low energies we  have the analytical
result \cite{Harmark:2004ws}
\begin{equation}
S_{\rm loc} (E) \simeq \frac{2 \pi^{1/3}}{3} (\frac{6}{5})^{5/6}
\frac{ \sqrt{N}}{(2\pi)^4} \frac{V_5}{l_s^5}
\left( (2\pi)^5 \frac{l_s^6 E}{V_5} \right) ^{5/6}
\left(1 + \frac{\zeta(3)}{10 \pi^2}
(2\pi)^5 \frac{l_s^6 E}{V_5} \right) \spa
 \frac{l_s^6 E}{V_5}
\ll 1  \ .
\end{equation}
This was computed using results of \cite{Harmark:2003yz}. The
entropy function for the entire localized phase, as displayed in
Fig.~\ref{figm5ent}, is computed here using the data of
\cite{Kudoh:2004hs} and the map of \cite{Harmark:2004ws}.
\end{itemize}

In the microcanonical ensemble, the phase with the highest entropy
dominates. From Fig.~\ref{figm5ent} we thus conclude that for
energies below $E_{\#}/V_5 = 2.51 \cdot (2\pi)^{-5} l_s^{-6}$,
 the localized phase is preferred in
the canonical ensemble, whereas for energies above that value the
system is in the uniform phase. In the range $E_{\#} < E \leq
E_{\rm max}$, with $E_{\rm max}/V_5 = 3.92 \cdot (2\pi)^{-5}
l_s^{-6}$, we have that the entropy of the uniform phase is
greater than that of the localized phase. For $E > E_{\rm max}$
only the uniform phase is available. Note that $ E_{\#} <
E_{\star} < E_{\rm max}$. The non-uniform phase always has lower
entropy than the two other phases.

Note that the point of the localized phase in which $E = E_{\rm
max}$ is special. Here we have zero heat capacity, which
corresponds to a singular behavior in the microcanonical ensemble.
If we imagine an infinitesimal change in the system with energy
change $dE$ and a corresponding change in the temperature $dT$, we
have that $dT / dE$ diverges as we approach the energy $E_{\rm
max}$. The energy $E_{\rm max}$ can therefore be seen as a
limiting energy for the localized phase in this sense.

It is interesting to consider what happens when throwing in a
finite amount of energy $\Delta E$, starting with an energy $E <
E_{\#}$. Three things can happen. If $E + \Delta E < E_{\#}$ we
stay in the localized phase. If $E_{\#} < E + \Delta E < E_{\rm
max} $ we also jump to a point in the localized phase. However,
contrary to before the entropy of the uniform phase at the energy
$E + \Delta E$ is greater so the system will go through a phase
transition and end up in the uniform phase. Finally, if $E +
\Delta E > E_{\rm max}$ the only available phase is the uniform
phase, so the system presumably has to end up in that phase.

\subsubsection*{Comparing the ensembles}

If we compare the canonical and microcanonical ensembles, we see
significant differences in the qualitative behavior of the
near-extremal NS5-brane thermodynamics. In the canonical ensemble,
the localized phase (in the interval given by $E \leq E_{\star}$)
is always the dominant phase, and it has a positive specific heat
and a limiting temperature $T_{\star}$.

Considering instead the microcanonical ensemble, we see that for
energies $E \leq E_{\#}$ the localized phase is again the dominant
phase, just as in the canonical ensemble. However, the qualitative
behavior is very different for energies $E_{\#}< E < E_{\rm max}$.
In that energy range the uniform phase is dominant, and hence we
have two different preferred phases in the two ensembles. For
energies $E > E_{\rm max}$ the difference is even more significant
since here only the uniform phase is available. For large
energies, it thus seems that the two ensembles are completely
inequivalent.

Since this system is without gravity one would expect that it
always is possible to bring the system in contact with a heat bath
of a certain temperature. Therefore, the inequivalence of
ensembles could suggest that there is a subtlety in the definition
of the canonical ensemble. Alternatively, it could point to the
existence of a new hitherto unknown phase of near-extremal
NS5-branes, also since this potentially could solve the problem
that we seemingly do not have any stable phases with temperatures
greater than $T_{\star}$. See the conclusions in Section
\ref{sec:con} for further discussion on these issues.

%%%%%%%%%%%%%%%%%%%%%%%%%%%%%%%%%%%%%%%%%%%%%%%%%%%%%
\section{Thermodynamics of D$p$-brane case}
\label{sec:Dp}

For comparison, we comment in this section on the thermodynamics
for near-extremal D$p$-branes on a transverse circle. Following
\cite{Harmark:2004ws}, this is related to the thermodynamics of phases of
Kaluza-Klein black holes in $10-p$ dimensions.

Our statements below concern D0, D1, D2 and D3-branes, which are
thus related to phases of Kaluza-Klein black holes in 10, 9, 8 and
7 dimensions. Just as in six dimensions, we know that there exists
a uniform black string, non-uniform black string and localized
black hole phase. However, contrary to the six-dimensional case,
the latter two branches are not known numerically. Nevertheless,
we are still able to make some general observations, based on the
results of \cite{Harmark:2004ws}.

From the map of \cite{Harmark:2004ws} these three phases of
neutral Kaluza-Klein black holes generate, just as for the
M5-brane case, three corresponding phases of near-extremal
D$p$-branes on a circle: The uniform, non-uniform and localized
phase. Moreover, the corresponding curves in the $(E,r)$ phase
diagram are believed to be qualitatively of the same form as in
that for the NS5-brane given in Fig.~\ref{figm5a}. However, the
thermodynamics is very different in nature, as we now discuss.

For definiteness, we focus here on $N$ near-extremal D2-branes on
a transverse circle, which is dual to finite temperature N=4
supersymmetric $SU(N)$ Yang-Mills on $\R^2\times \T^2$ (the $\T^2$
consists of a compact spatial direction of circumference $\hat L$
together with the compact Euclidean time direction). The
qualitative picture is expected to hold for the other D$p$-brane
cases as well. The difference with the M5-brane case is best
illustrated by looking at the $T(E)$ diagram.

For the uniform phase, consisting of near-extremal D2-branes
uniformly smeared on a transverse circle, we have the dependence
\begin{equation}
T(E) \propto E^{\frac{1}{4}} \ ,
\end{equation}
i.e. the thermodynamics of near-extremal D3-branes. The uniform
phase dominates at high energies. For the localized branch which
dominates at low energies we know from \cite{Harmark:2004ws} that
\begin{equation}
T(E) \propto E^{\frac{3}{10}} ( 1 + \alpha E )  \spa E \ll 1 \ ,
\end{equation}
with $\alpha < 0$ a known constant. Finally, for the non-uniform
phase, we know that it emanates from the uniform phase at the
critical energy $E_{\rm c}=  0.71 \cdot (2\pi)^3 V_2 N^2/(\lambda
\hat L^3)  $. At this critical point we know the ratio of the
specific heat of the non-uniform phase and the uniform phase,
which is given by \cite{Harmark:2004ws}
\begin{equation}
\label{cnu} \frac{c_{\rm nu}}{c_{\rm u}} = \frac{1}{1 +
\frac{6}{7} \hat \gamma \epsilon_c} = 0.41 \  .
\end{equation}
In particular the specific heat in that point is smaller than that
of the uniform phase, a feature which holds for any of the D$p$-brane
cases mentioned above.

\begin{figure}[ht]
\centerline{ \epsfig{file=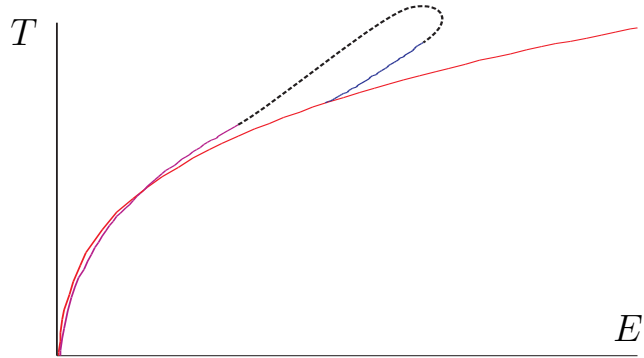,width=8 cm,height=5cm} }
\caption{Qualitative picture of energy versus temperature for the
phases of near-extremal D$p$-branes on a transverse circle. Shown
are the uniform (red), non-uniform (blue) and localized (magenta)
phases. The dashed curve connects the localized and non-uniform
phase. \label{figdp}}
\begin{picture}(0,0)(0,0)
\put(87,215){\Large $T$} \put(315,103){\Large $E$}
\end{picture}
\end{figure}

Based on these results, we have the following features of
the three phases in a $T(E)$ diagram:
\begin{itemize}
\item The uniform and localized phase start in the origin (with infinite slope)
and for very low energies the uniform phase has higher temperature than
the localized phase at a given energy. We expect that the localized phase will
then cross the uniform phase already for relatively low energies,
at least below the critical energy where the non-uniform phase starts.
\item Since the slope in the $T(E)$ diagram is the inverse of the specific heat,
it follows from \eqref{cnu} that the slope of the non-uniform phase at the
critical point $(E,T)=(E_{\rm c}, T_{\rm c})$ is larger than that of
the uniform phase.
\item We assume that the localized and non-uniform branch meet in a
qualitatively similar way, as observed explicitly in Fig.~\ref{figm5te} for
the NS5-brane.
\end{itemize}
We are then led to a picture that is summarized in the $T(E)$ diagram of
Fig.~\ref{figdp}.

We note that the picture of the thermodynamics displayed in Fig.~\ref{figdp}
implies that the $(E,r)$ phase diagram for near-extremal D$p$-branes on
a transverse circle is qualitatively of the same form
as that for the NS5-brane in Fig.~\ref{figm5a}. This means in particular
that as the energy increases from zero in the localized phase we have that:
i) First the point is reached where the specific heat becomes infinite. Here
the maximal temperature of the localized phase occurs; ii) Increasing the energy further
one reaches the point where the specific heat is zero. Here the maximal energy
of this phase occurs.%
\footnote{The conditions for infinite and zero specific heat are
 in this case $E r'(E)= (5-p)/2 - r(E)$ and $1/r'(E) =0 $ respectively.}

Hence, just as in the NS5-brane case, the localized phase has a maximal
temperature. On the other hand, contrary to NS5-brane case,
where the uniform phase has the constant Hagedorn temperature, the uniform
phase exists for any temperature.
As a consequence there is no limiting temperature in the system when
 considering near-extremal D$p$-branes on a circle.
  In particular, for sufficiently high temperatures
the system will be in the uniform phase. Below some critical temperature
the localized phase will be favored. Moreover, the canonical and microcanonical
ensemble show the same behavior.

%%%%%%%%%%%%%%%%%%%%%%%%%%%%%%%%%%%%%%%%%%%%%%%%%
\section{Conclusions}
\label{sec:con}

We have examined the thermodynamics of the near-extremal type IIA
NS5-brane using its description as an M5-brane with a transverse
circle, and the relation of the latter to the phases of
six-dimensional Kaluza-Klein black holes. We have found that in
the canonical ensemble there is a new background of near-extremal
M5-branes localized on the transverse circle, which dominates over
the usual near-extremal NS5-brane background. This new phase has
the surprising feature that it has a limiting temperature
$T_{\star} = 1.177 \cdot T_{\rm hg}$, which lies above the
Hagedorn temperature $T_{\rm hg}$ of the usual NS5-brane phase. We
have also considered the microcanonical ensemble and compared it
with the canonical ensemble, indicating some significant
differences between the two.

\begin{figure}[ht]
\centerline{\epsfig{file=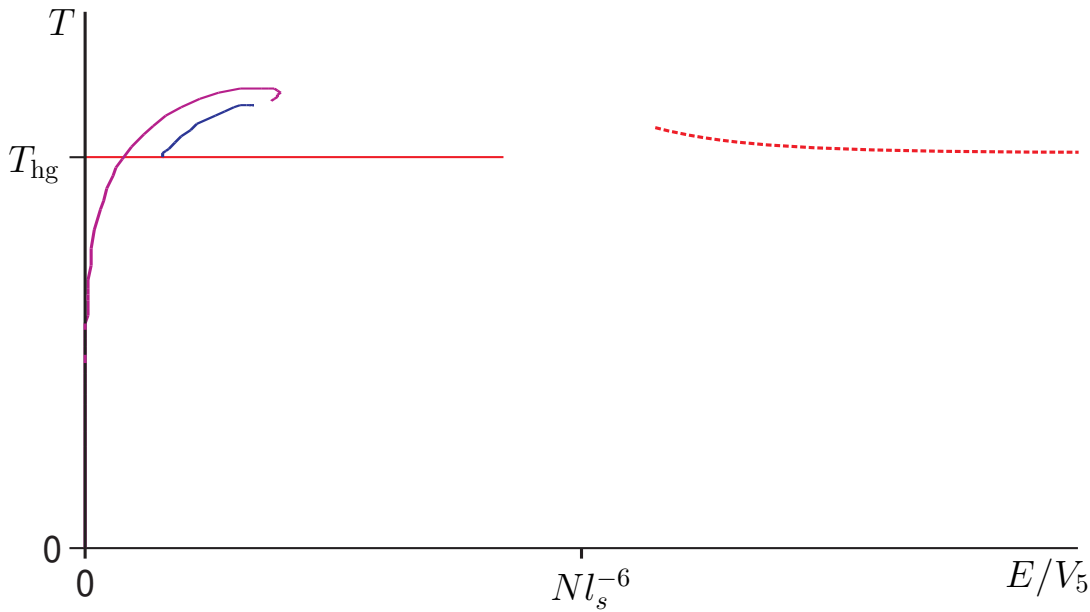,width=14cm,height=8cm} }
\caption{Temperature versus energy for the near-extremal type IIA
NS5-brane. Displayed are the same phases as in Figure
\ref{figm5te}, but in addition we have sketched the high energy
phase of the near-extremal NS5-brane. \label{figHE} }
\begin{picture}(0,0)(0,0)
\put(18,289){\Large $T$} \put(3,237){\Large $T_{\rm hg}$}
\put(208,75){\Large $N l_s^{-6}$} \put(380,80){\Large $E/V_5$}
\end{picture}
\end{figure}

We conclude with the following remarks:

{\bf High energy regime versus low energy regime.} The analysis of
the type IIA NS5-brane thermodynamics of this paper is valid in
the low energy regime $E \ll N V_5 l_s^{-6}$. This is the regime
in which the effective string coupling is sufficiently large to
see the non-perturbative effects of the eleven-dimensional circle.
In the high energy regime $E \gg N V_5 l_s^{-6}$ we instead have
that the effective string coupling is small. In fact, the
near-extremal NS5-brane background has an exact worldsheet CFT
description which is asymptotically free in the sense that we can
define a perturbation theory around infinite energy in powers of
$1/E$ \cite{Kutasov:2001uf}. The results of \cite{Kutasov:2000jp}
then suggest that the near-extremal type IIA NS5-brane has a
high-energy phase that has $T> T_{\rm hg}$ and negative specific
heat. In figure \ref{figHE} we have made a sketch of this phase in
an $E$ versus $T$ diagram (see \cite{Parnachev:2005hh} for a
similar sketch) together with the phases described in this paper
as
displayed in Figure \ref{figm5te}.%
\footnote{According to
Refs.~\cite{Rangamani:2001ir,Buchel:2001dg} the instability of the
high energy thermodynamics of near-extremal NS5-branes found in
\cite{Kutasov:2000jp} should be holographically dual to a
Gregory-Laflamme like instability, corresponding to longitudinal
perturbations of the NS5-brane. This is in contrast to
Ref.~\cite{Kutasov:2000jp} where it is argued that the instability
is of stringy origin.}

Now, the results of this paper are all concerned with the low
energy regime $E \ll N V_5 l_s^{-6}$. This is the regime in which
the effective coupling of the exact worldsheet CFT description for
the type IIA NS5-brane background is strong. We see thus that in
the microcanonical ensemble there are no discrepancies or problems
with the existence of the new phases of this paper, since we have
a clear distinction between the low energy and high energy
regimes. The problems only appear when going to the canonical
ensemble. If we start at very low temperatures and heat up the
type IIA near-extremal NS5-brane, we will be in the localized
phase all the way up to the temperature $T_\star$, which lies
above $T_{\rm hg}$. Here the canonical ensemble seems
well-defined, as the localized phase of near-extremal M5-branes
clearly does not have large fluctuations around it. On the other
hand, seen from the point of view of the high energy regime, it is
possible that the fluctuations of the energy are large which seems
to suggest that the canonical ensemble is ill-defined near the
temperature $T_{\rm hg}$. We thus have an apparent discrepancy.
However, in our opinion this discrepancy cannot mean that the
canonical ensemble is ill-defined as one heats up the system from
zero temperature to $T_\star$. This is because such a process is
clearly described by classical solutions and we know that
fluctuations around these solutions are suppressed, which means
that we have a well-defined temperature. Hence, the discrepancy
must have another source. We think that this should be very
interesting to examine further.

{\bf Near-extremal NS5-brane phases with bubbles}. In this paper,
we have focussed on the type IIA NS5-brane phases that can be
obtained from six-dimensional Kaluza-Klein black holes without
Kaluza-Klein bubbles. However, there exists a large class of
six-dimensional bubble-black hole sequences \cite{Elvang:2004iz}
that contain both event horizons and Kaluza-Klein bubbles. These
can be mapped to corresponding phases of near-extremal NS5-branes
involving Kaluza-Klein bubbles \cite{Harmark:2004bb}. For the
purpose of this paper, these are not relevant as they are
subdominant in both the canonical and the microcanonical ensemble.
In particular, in the canonical ensemble one finds that the
NS5-brane/bubble phases have positive free energy, while we also
note that the temperatures of these phases approach the Hagedorn
temperature of the uniform phase in the limit where the size of
the bubble goes to zero. In the microcanonical ensemble one
observes that the entropy of the NS5-brane/bubble phases lies
below that of the uniform phase, and approaches the Hagedorn
behavior $S \simeq E/T_{\rm hg}$ of the uniform phase in the limit
where the size of the bubble goes to zero. Finally, it is amusing
to note that in many of the phases involving Kaluza-Klein bubbles
one can have arbitrarily high temperatures.

{\bf Putting fluxes on the type IIA NS5-brane}. One can deform the
near-extremal NS5-brane by adding a D0, D2 or D4-brane flux on the
world-volume. The dual theories on the world-volume of the
NS5-brane are the OD-theories
\cite{Harmark:2000ff,Gopakumar:2000ep}. These should be possible to
study using the techniques of this paper. This could be
interesting in view of the recently studied gravitational and
thermodynamic instabilities \cite{Friess:2005tz,Ross:2005vh} of
these types of brane bound states.

{\bf Type IIB NS5-brane}. An obvious question to address is if our
results of this paper on the thermodynamics of the type IIA
NS5-brane also can be used to say anything about the type IIB
NS5-brane, since these are related by a longitudinal T-duality.
However, if we want to take a decompactification limit of the
T-dual circle on the IIB side, it corresponds to a circle of
string size at the IIA side, and thus we cannot use supergravity
on the IIA side to predict the behavior on the IIB side. In
particular, the new localized phase for the type IIA NS5-brane
obtained in this paper cannot be related to a corresponding phase
for the (decompactified) type IIB NS5-brane. Therefore, it seems
that a separate study is required for the thermodynamics of the
near-extremal type IIB NS5-brane. Another way to understand this
is to notice that the low energy behavior of the type IIB
near-extremal NS5-brane is given by a six-dimensional
super-Yang-Mills theory. Therefore, one would have to understand
the low energy behavior using gauge theory rather than gravity.

\section{Acknowledgments}

We thank Micha Berkooz, Jan de Boer, Poul Henrik Damgaard, Vasilis
Niarchos and Peter Orland for useful discussions. Special thanks
to Hideaki Kudoh and Toby Wiseman  for explanations and providing
us with the numerical results of Ref.~\cite{Kudoh:2004hs}. Work
partially supported by the European Community's Human Potential
Programme under contract MRTN-CT-2004-005104 `Constituents,
fundamental forces and symmetries of the universe'.

%\addcontentsline{toc}{section}{References}

%The following two lines is for bibtex only:
%\bibliographystyle{utphys}
%\bibliography{bibrot,biblioniels}
%\bibliographystyle{../INPUT/utphys}
%\bibliography{../BIB/bibrot,../BIB/biblioniels}
%\bibliographystyle{C:/mytex/INPUT/utphys}
%\bibliography{C:/mytex/BIB/bibrot,C:/mytex/BIB/biblioniels}

\providecommand{\href}[2]{#2}\begingroup\raggedright\endgroup

\end{document}